# Synthesis and superconducting properties of the iron oxyarsenide TbFeAsO$_{0.85}$

Y.G. Shi,[1,*] S. Yu,[1] A.A. Belik,[2] Y. Matsushita,[3] M. Tanaka,[3] Y. Katsuya,[4] K. Kobayashi,[3] K. Yamaura,[1] E. Takayama-Muromachi[1,2]

[1] Superconducting Materials Center, National Institute for Materials Science, 1-1 Namiki, Tsukuba, 305-0044 Ibaraki, Japan
[2] International Center for Materials Nanoarchitectonics, National Institute for Materials Science, 1-1 Namiki, Tsukuba, 305-0044 Ibaraki, Japan
[3] NIMS Beamline Station at SPring-8, National Institute for Materials Science, 1-1-1 Kouto, Sayo-cho, Sayo-gun, Hyogo 679-5148, Japan
[4] SPring-8 Service Co. Ltd., 1-1-1 Kouto, Sayo-cho, Sayo-gun, Hyogo 679-5148, Japan

**Abstract**

The iron oxyarsenide TbFeAsO$_{0.85}$ was synthesized by a high-pressure method. A high-quality polycrystalline sample was obtained and characterized by measurements of magnetic susceptibility and electrical resistivity. Bulk superconductivity with $T_c$ of 42 K was clearly established without an F doping usually conducted to tune on superconductivity in the iron oxypnictide.

---

[*] E-mail address: SHI.Youguo@nims.go.jp

**Introduction**

The iron oxypnictide superconductor was an epoch discovery in research of high-$T_c$ superconductivity because it directly proved that the copper oxide is no longer a unique substance showing high-$T_c$ superconductivity. After several initial reports [1,2], large number of submissions about the superconductor appeared in the preprint server [3], directly indicating that enormous attentions focused into the new subject. To date, within our best knowledge, the highest $T_c$ was reported 56 K for SmFeAs(O,F) [4], being truly remarkable.

Many possibilities have been suggested so far regarding future applications and technological merits. Besides, new insights into underling physics of high-$T_c$ superconductivity are highly expected to come out because the iron oxypnictide may share common physics with the copper oxide superconductor. However, it becomes obvious that further materials developments are necessary to put forward present researches, since many experimental possibilities still remains untested. In addition, synthesis of the superconductor is highly complicated in nature. Thus, further synthesis efforts are extensively required.

In series of the F doped iron oxyarsenide $Ln$FeAs(O,F) ($Ln$: rare earth element), high quality polycrystalline samples with $Ln$ = La, Pr, Nd, and Sm were already prepared, while, with smaller $Ln$ ($\leq$ Eu) the synthesis was achieved within certain limits. Recently, we indeed achieved a high-quality synthesis with $Ln$ = Tb by using a high-pressure method, followed by primary physical properties characterizations. In this paper, we report synthesis and superconducting properties of the newly achieved iron oxyarsenide TbFeAsO$_{0.85}$, in which F was un-doped and oxygen vacancies were instead introduced to turn on superconductivity.

**Experimental**

Polycrystalline TbFeAsO$_{0.85}$ was prepared by a solid-state-reaction method. At first, a precursor TbAs was prepared from Tb (20 mesh 3N in oil, Rare Metallic Co.) and As (powder 5N, High Purity Chemicals) with excess 1 mole % of As (i.e. starting composition Tb/ As = 1/ 1.01). Beforehand, the oil was carefully removed by using CCl$_4$ in an ultra sonic bath. A mixture of the sources was put into an evacuated quartz tube and then slowly heated up in a furnace. The heating was held at 500 °C for 10 hrs, followed by cooling down. The product was carefully ground with an agate mortar and a pistol by hand, and then heated again at 850 °C for 8 hrs in the same manner.

A weighted powder Tb/ Fe/ As/ O= 1/ 1/ 1/ 0.85 was prepared from TbAs, α-Fe$_2$O$_3$ (0.46μm 3N, Furuuchi Chemical Co.), and Fe (100 mesh 3N, Rare Metallic Co.), and it was placed into a Ta capsule with a h-BN inner. The BN was pre-heated at 1600 °C for few hours in vacuum to remove small amount of possible boric oxide. The loaded capsule was heated in a high-pressure apparatus, which was capable of maintain 6 GPa during heating. The elevated temperature was 1450 °C and the heating duration was 3 hrs. The capsule was then cooled down to room temperature within few minutes before releasing the pressure. The pellet was ground and heated again in the high-pressure furnace at 1450 °C for 4 hrs at 6 GPa in the same manner.

The obtained pellet, dense and black in color, was investigated by a powder X-ray diffraction (XRD) method in a commercial apparatus (Rigaku, RINT2200V/PC). Cu-Kα radiation was used. The 2 theta range was between 5° and 100°.

Electrical resistivity of the pellet was measured by a four-points probe method with ac gage current of 1 mA at 30 Hz in a commercial apparatus (Quantum Design,

Physical Properties Measurement System). Electrical contacts on the four locations along the pellet were prepared from gold wires and silver paste. Magnetic susceptibility was measured after cooling the pellet to 2 K without applying a magnetic field (zero-field cooling, ZFC), then slowly warmed up to 300 K and then cooled down to 2 K in a magnetic field of 10 Oe (field cooling, FC) in a commercial magnetometer (Quantum Design, Magnetic Properties Measurement System).

**Results and discussion**

Fig. 1 shows an XRD profile of TbFeAsO$_{0.85}$ measured at room temperature. As marked by *hkl* indexes, all major peaks were clearly characterized by assuming a tetragonal unit cell with $a$ = 3.889(1) Å and $b$ = 8.376(1) Å. The unit cell size was comparable with what was reported elsewhere for the F-doped compound TbFeAs(O,F) [5]. It should be noted that a small amount of TbAs was found in the analysis (as marked by stars in the pattern).

We actually tested a role of the excess As (0.01 mole per TbAs) in the precursor preparation, and found that it was crucial for the quality of the final product. The XRD quality of a final product prepared without adding the excess As was fairly low. Much excess As (i.e. TbAs$_{1.02}$) resulted in visible As inside of the quartz tube after heating, suggesting that the quantity 0.01 mole of As was nearly optimized.

Fig. 2 (top panel) shows temperature dependence of the electrical resistivity of the polycrystalline sample of TbFeAsO$_{0.85}$, showing clearly a superconducting transition at 42.0 K. The transition width is relatively sharp <3 K (see the expansion in the figure), suggesting a high quality of the sample. A corresponding drop of the magnetic susceptibility was observed at 41.5 K with a large superconducting volume fraction: the shielding fraction (ZFC data) was 1.12 at the low temperature limit (1.00 for the perfect diamagnetism). The calculated density 7.95 g/cm$^3$ was used in the estimation. Both the electrical and the magnetic data clearly established a bulk superconductivity of TbFeAsO$_{0.85}$.

On the other hand, the Meissner volume fraction (FC data) was smaller <0.1 than what was reported elsewhere for the comparable $T_c$ superconductor SmFeAsO$_{1-x}$F$_x$ ($T_c$ = 43 K) [6]. It is likely due to robust magnetic flux in the superconducting state, suggesting that the oxygen vacancies may work as a pining center much effectively rather than the doped F elements.

As mentioned, the magnetic and electrical properties accords exactly with a bulk superconductivity, however a small step was seen at ~30 K in the ZFC curve, suggesting a possible chemical inhomogeneity in the sample. Further heating under the high pressure condition could address the issue.

The bulk superconductivity for TbFeAsO$_{0.85}$ was eventually observed after the second heating in the high-pressure furnace. Superconducting properties observed after the first heating were entirely poor and weak: a much small superconducting volume fraction and lower $T_C$ of ~20 K, reflecting that solid-state reactions run much slowly even under the high pressure condition. The XRD analysis indeed found multiple compounds including TbAs and Fe$_2$O$_3$ in the sample heated just for once. We found the poor situation was dramatically improved by repeating the heating in the high-pressure furnace.

Within our attempts, synthesis of TbFeAsO$_{0.85}$ without a high pressure method was unachieved because it required a high-temperature heating beyond the quartz tube limit. It was, thus, clear that the high-pressure method was essential to prepare the high quality polycrystalline TbFeAsO$_{0.85}$.

The oxygen quantity was fixed at 0.85 per the mole because an optimized superconductivity was claimed at the concentration in the *Ln* = La to Sm studies [7]. In this study, we tried to synthesize the oxygen stoichiometric TbFeAsO under the same condition, however, the XRD sample quality remained very low, suggesting that a state with the oxygen vacancies is rather stable than a state without those in the Tb system. The introduced vacancies may form a super-lattice cell, resulting in the stability. Thus, it could be useful to investigate the samples in an electron microscopy to judge the possibility.

A recent synthesis using a high pressure method indicates that the F-doped compound TbFeAsO$_{0.9}$F$_{0.1}$ turns into a superconducting state at 45-46 K as well as TbFeAsO$_{0.85}$ [5]. Let us compare the formal doping level: doped F$_{0.1}$ causes 0.1 electrons per the mole, while the oxygen vacancies (TbFeAsO$_{0.85}$) do 0.3 electrons per the mole. The estimation suggests a possibility that TbFeAsO$_{0.85}$ is over doped with respect to the charge carrier concentration. Thus, reducing the oxygen-vacancy concentration possibly enhances $T_c$: further studies are in progress.

**Acknowledgments**

This research was supported in part by the Superconducting Materials Research Project from MEXT, Japan, the Grants-in-Aid for Scientific Research from JSPS (18655080,20360012), the Murata Science Foundation, and the Futaba Memorial Foundation.

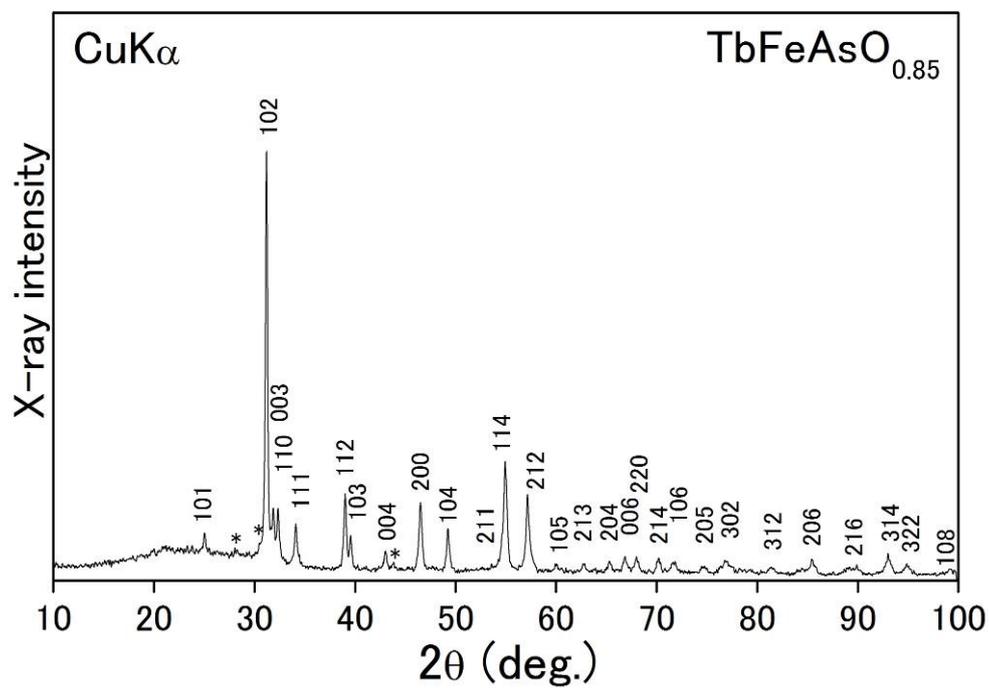

Fig. 1 Powder X-ray diffraction profile of the superconducting compound TbFeAsO$_{0.85}$. A small amount of TbAs was detected as marked by stars.

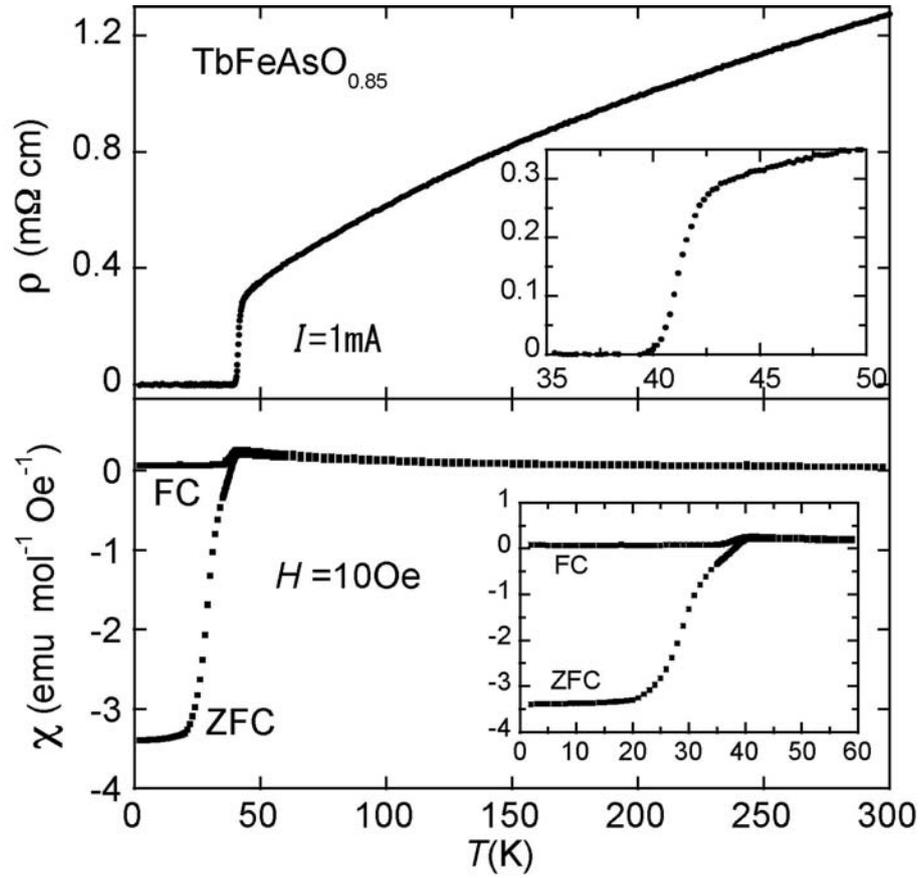

Fig. 2 Electrical resistivity and magnetic susceptibility of the polycrystalline compound TbFeAsO$_{0.85}$. Inset is an expansion of each data.